\documentclass[12pt]{article}
\input{epsf.tex}
\usepackage[english]{babel}
\usepackage{graphics}
\usepackage{color,rotating}
\usepackage{pictex}
\usepackage{latexsym,amsfonts,amssymb}
\newcommand{\beq}{\begin{equation}}
\newcommand{\eeq}{\end{equation}}
\newcommand{\bea}{\begin{eqnarray}}
\newcommand{\eea}{\end{eqnarray}}
\newcommand{\bdm}{\begin{displaymath}}
\newcommand{\edm}{\end{displaymath}}
\newcommand{\ba}{\begin{array}}
\newcommand{\ea}{\end{array}}
\newcommand{\ben}{\begin{enumerate}}
\newcommand{\een}{\end{enumerate}}
\newcommand{\bde}{\begin{description}}
\newcommand{\ede}{\end{description}}

\newcommand{\ci}{{\ \hbox{{\rm I}\kern-.6em\hbox{\bf C}}}}%simbolo di numero complesso%
\newcommand{\erre}{{\hbox{{\rm I}\kern-.2em\hbox{\rm R}}}}%simbolo di numero reale%
\newcommand{\1}{ \,  \raisebox{+0.14em}{{\hbox{{\rm \scriptsize ]}} \raisebox{-0.2em}{\kern-.8em\hbox{1}}}} \, }
\newcommand{\Z}{{\bf Z} }%simbolo di numero razionale%

\renewcommand{\a}{\alpha}
\renewcommand{\b}{\beta}

\renewcommand{\d}{\delta}

\newcommand{\ka}{\kappa}

\begin{document}

\begin{flushright}
% put a pre-print number in the top-right
hep-th/yymmnnn\\
% and today's date
% \today
\end{flushright}

\begin{center}
\Large
{\bf Low-energy effective theory for a Randall-Sundrum scenario with a moving bulk brane}
\end{center}

\begin{center}
Ludovica Cotta-Ramusino$^{*,\dagger}$ and David Wands$^*$ \\
{\textit{$^*$Institute of Cosmology and Gravitation, University of Portsmouth, Portsmouth, P01 2EG, United Kingdom}}\\
{\textit{$^\dagger$Laboratory for Computation and Visualization in Mathematics and Mechanics, EPFL FSB IMB, Ecole Polytechnique
F\'ed\'erale de Lausanne, CH-1015, Switzerland.}}
\end{center}

\vskip0.2in

\begin{abstract}
We derive the low-energy effective theory of gravity for a generalized Randall-Sundrum scenario,
allowing for a third self-gravitating brane to live in the 5D bulk spacetime. At zero order the 5D spacetime is composed of
two slices of anti-de Sitter spacetime, each with a different curvature scale, and the 5D Weyl tensor vanishes. Two
boundary branes are at the fixed points of the orbifold whereas the third brane is free to move in the bulk.
At first order, the third brane breaks the otherwise continuous evolution of
the projection of the Weyl tensor normal to the branes. We derive a junction condition
for the projected Weyl tensor across the bulk brane, and
combining this constraint with the junction condition for the extrinsic curvature tensor, allows us to
derive the first-order field equations on the middle brane. The effective theory is a
generalized Brans-Dicke theory with two scalar
fields. This is conformally equivalent to Einstein gravity and two scalar fields, minimally coupled
to the geometry, but nonminimally coupled to matter on the three branes.
\end{abstract}

\section{Introduction}
\setcounter{equation}{0}

The possibility that our Universe might be a $(3+1)$-dimensional
membrane (brane) embedded in some higher $(4+n)$-dimensional
spacetime (bulk), as suggested by M-theory \cite{M theory}, has
been extensively studied over recent years. In brane models,
although gravity can propagate in the whole bulk, other matter
fields are localized on the brane.
In particular, Randall and Sundrum \cite{Randall
SundrumI,Randall SundrumII} proposed a brane-world model in which
we live on a 4D brane embedded in a $5D$ anti-de Sitter (AdS)
spacetime. In \cite{Randall SundrumI} two branes are placed at the
fixed points ($y=0$ and $y=y_0$) of an orbifold $\bf{S}^1/\Z_2$,
where we identify $y \rightarrow -y$ and $y-y_0 \rightarrow
y_0-y$, and $y$ is the extra-coordinate. The bulk AdS spacetime
can be then thought of as being bounded by the two $4D$ branes.
The induced metrics on the branes can be flat (Minkowski) if a
fine tuning condition is imposed on the vacuum energies or
tensions of the branes:
 \beq
  \sigma_0 = -\sigma_{y_0}=
\frac{6}{\ka_5^2}\frac{ 1}{\ell}
 \label{fine tuningRS}.
 \eeq
where $\sigma_0$ and $\sigma_{y_0}$ are the tensions on the hidden
brane at $y=0$ at on the visible brane at $y=y_0$, $\ka_5^2$
is the ($5D$) gravitational coupling and
$\ell$ is the AdS curvature scale.

Assuming the bulk metric obeys the 5D vacuum Einstein equations,
then the projected gravitational field equations on the brane are
modified with respect to general relativity \cite{Giappo}. Two
additional terms appear with respect to general relativity: a
local term, quadratic in the energy-momentum tensor on the brane,
and a non-local term, which is a projection of the $5D$ Weyl
tensor, namely $E_{\a\b}$.

The contribution to the $4D$ effective theory of $E_{\a\b}$, which
describes the contribution of the bulk gravitational field on the
brane and influences the brane cosmological evolution, is of
crucial importance \cite{Maartens}. Although the
quadratic source term becomes relevant only at high energies, the
projected Weyl tensor may remain non-negligible in the low-energy
regime, where one would hope to recover general relativity.

The projected Weyl tensor generally has non-closed equations on
the brane \cite{Giappo,Maartens} and in general one should solve
the full bulk gravitational field equations.
However it is possible to derive a scheme which allows one to
self-consistently solve the $5D$ Einstein equations in the 
low-energy regime, and carefully construct the projected Weyl tensor
on the brane.

A low-energy perturbation scheme was proposed in \cite{AdS/CFT}
for the Randall-Sundrum (RS) two brane scenario \cite{Randall
SundrumI}. (See also \cite{Wiseman}.) The low-energy regime is defined as the
regime in which the matter energy density on the brane is much
smaller than the RS brane tension (\ref{fine tuningRS}). The
perturbation parameter is defined as the ratio between these two
energy densities, and the $5D$ Einstein equations can be solved at
different orders in the perturbation parameter. This method allows
one in principle to derive the effective Einstein equations on the
brane at each order, although of particular interest is the first
order
correction, which is the most relevant at low energies.

In the original derivation by Kanno and Soda \cite{AdS/CFT}, the
full $5D$ equations of motion were solved at each order in the
bulk by performing a perturbation expansion in the metric. In the
alternative derivation by Shiromizu and Koyama \cite{Low energy},
the expansion was rather done directly in terms of the extrinsic
curvature and the projected Weyl tensor, whose equations of motion
can then be solved in the bulk at each order. Crucially, the use
of the junction conditions enables one to express the Weyl tensor
as a function of the matter content of the branes and the physical
distance between the branes, interpreted as the radion field. This
finally allows for the derivation of the effective Einstein
equations induced on the brane at low energy. On both branes, at
first order, the final effective theory is called by the authors
of \cite{AdS/CFT,Low energy} a quasi-scalar-tensor gravity theory,
where the Brans-Dicke field couples through different
gravitational coupling constants with matter on each brane.

The low-energy effective actions on the negative and positive
brane respectively are \cite{AdS/CFT,Low energyII}
 \bea
 S_{-} & = & \frac{\ell}{2\ka_5^2} \int{d^4x\sqrt{-g}\left[\Phi R -
\frac{\omega_-(\Phi)}{\Phi}(\nabla\Phi)^2\right]} \nonumber \\
& + &
\int{d^4x\sqrt{-g}\left(\mathcal{L}_{-} +(1+\Phi)^2
\mathcal{L}_{+}\right)} \label{effective action neg brane}
 \eea
where $\omega_-(\Phi) = -3\Phi[2(\Phi +1)]^{-1}$, and
 \bea
S_{+} & = & \frac{\ell}{2\ka_5^2} \int{d^4x\sqrt{-\hat{g}}\left[\Psi \hat{R} -
\frac{\omega_+(\Psi)}{\Psi}(\hat{\nabla}\Psi)^2\right]} \nonumber \\
& + &
\int{d^4x\sqrt{-\hat{g}}\left(\mathcal{L}_{+} +(1-\Psi)^2
\mathcal{L}_{-}\right)} \label{effective action pos brane}
 \eea
where $\omega_+(\Psi) = 3\Psi[2(1- \Psi)]^{-1}$, and ${\cal L}_-$ and ${\cal L}_+$ are the Lagrange densities for matter fields on the branes.

The theories, as expected for Brans-Dicke theories, are
conformally equivalent to Einstein gravity plus a
minimally-coupled scalar field, described by
 \bea
S_{EF} & = & \frac{\ell}{2 \ka_5^2}\int{d^4x \sqrt{-\tilde{g}}\left[\tilde{R}
-(\tilde{\nabla}{\chi})^2
\right]}\nonumber \\
& + &\int{d^4x \sqrt{-\tilde{g}}\,\,\left[\sinh^4{\left(\chi/\sqrt{6}\right)}\,\mathcal{L_{-}} +
\cosh^4{\left(\chi/\sqrt{6}\right)}\,\mathcal{L_{+}}\right]} \label{EFrame}
 \eea
where the conformal factors $\Omega^2$ (negative brane) and $\hat{\Omega}^2$ (positive brane) are
 \beq
\Omega^2 = \frac{1}{\sinh^2{(\chi/\sqrt{6})}}\hspace{1in}\hat{\Omega}^2 = \frac{1}{\cosh^2{(\chi/\sqrt{6})}}
 \eeq
and $\sqrt{(1+ \Phi)} = \coth{\left(\chi/\sqrt{6}\right)}$ and
$\sqrt{(1- \Psi)} = \tanh{\left(\chi/\sqrt{6}\right)}$, where $\Psi<1$ \cite{Turok, Born again, MPhil}.

There are many additional complications to take into account if
one wishes to relate simple brane-world models to realistic
configurations in the context of superstring and M-theory.
In any of brane-world models it is important to derive the
low-energy effective theory on the $4D$ branes.
In this paper we consider just the effect of adding an additional
brane in the bulk.
We will focus on a generalization of \cite{Low energy} to a three
brane scenario. The work presented in the paper was originally
presented in \cite{MPhil}. Since then a number of other authors
\cite{KKKK,KSW,deRham} have investigated the low-energy
effective theory corresponding to three branes in an AdS bulk, in
particular considering the effective potential for D-branes in
warped flux compactifications \cite{KKKK} and a simple geometrical
model for brane inflation \cite{KSW}.
A generalized Randall-Sundrum scenario with three branes was also
previously studied in \cite{Turok,Tre brane}, and multi-brane
collisions were considered in \cite{Collision of branes}.

The plan of the paper is the following. In the second section we
derive the low-energy effective theory in a generalized Randall-
Sundrum two-branes scenario, allowing for a third brane to live in
the bulk.
In section (\ref{preliminaries}) we discuss the preliminaries,
adapting the covariant formalism of \cite{Low energy} to a three
brane model. In section (\ref{zerorder}) we discuss the background
solution and (\ref{firstorder}) we derive the effective Einstein
equations at first order on the third brane. The effective theory
turns out to be a Brans-Dicke theory, with two independent scalar
fields, one minimally coupled with the geometry. Finally in
section (\ref{Eframe}) we show that a conformal transformation
relates the effective theory on the third brane to Einstein
gravity plus two minimally coupled scalar fields.
In the third and final section we draw our conclusions.
%%%%%%%%%%%%%%%%%%%%%%%%%%%%%%%%%%%%

%%%%%%%%%%%%%%%%%%%%%%%%%%%%%%%%%%%%
\section{First order effective theory at the third brane}\label{effectivetheory}

In this section we derive the low-energy effective theory at first
order in a three brane scenario, using the perturbative scheme
introduced in Ref.~\cite{AdS/CFT}. We choose to follow the
covariant approach as in Ref.~\cite{Low energy}.
%%%%%%%%%%%%%%%%%%%%%%%%%%%%%%%%
\subsection{Preliminaries} \label{preliminaries}

\begin{figure}
\begin{center}
\includegraphics[]{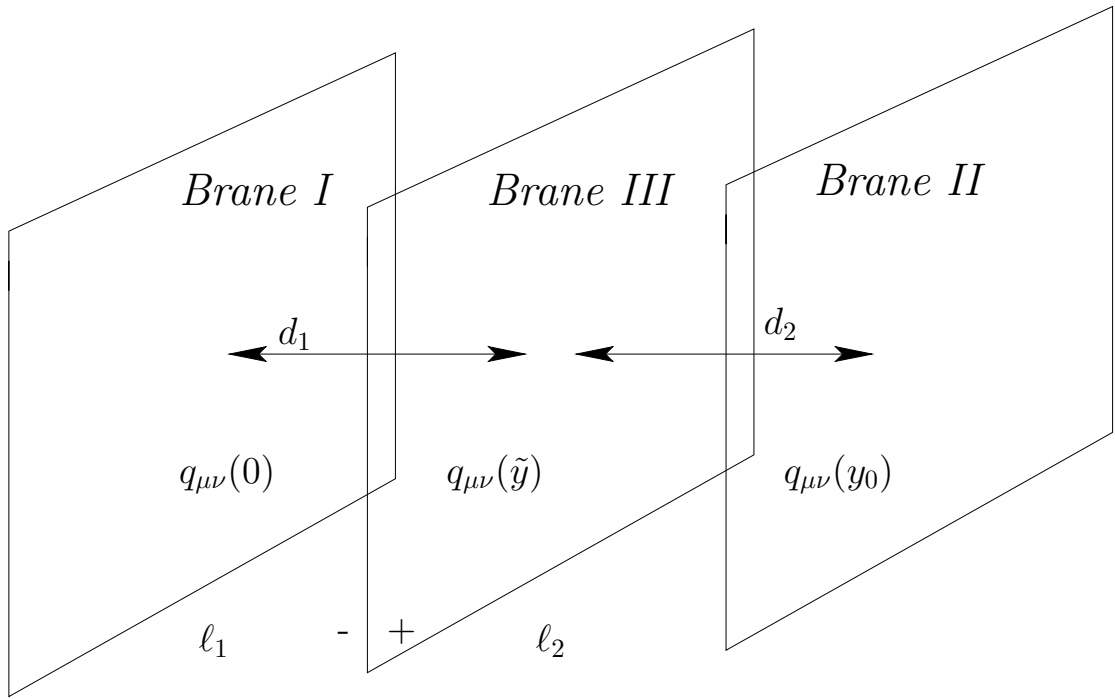}
\end{center}
\caption{The branes are separated by slices of AdS $5D$ spaces,
each with a different curvature scale, $\ell$. The sign $+$ and
$-$ are placed either side of the third brane to show our
convention for the orientation of the normal vector field. The
induced metrics, $q_{\mu\nu}$, on the three branes are conformally
rescaled by the geometrical warp factor.} \label{fig:brane_3_2}
\end{figure}

We consider an extension of the Randall-Sundrum two brane model
\cite{Randall SundrumI}. The three branes are separated by slices
of AdS bulk, each characterized by a different curvature scale, as
is schematically shown in figure~\ref{fig:brane_3_2}. Brane I and
II are still at the fixed points of the $\bf{S}_1/\bf{Z}_2$
orbifold, and therefore they respect a $Z_2$ symmetry, wheareas
Brane III is not at a fixed point of the orbifold. Two natural scalar
degrees of freedom characterize this scenario: one associated with
the overall distance between the two boundary branes, one
associated with the relative position of the third bulk brane.

The Randall-Sundrum metric \cite{Randall SundrumI} reads
 \beq
  ds^2 = e^{2\phi(y,x)}dy^2 + q_{\alpha\beta}(y,x)dx^{\alpha}dx^{\beta}
 \,,
\label{RS metric}
 \eeq
where $q_{\alpha\beta}$ is the metric
induced on the brane. The normal vector field $n^{\alpha}$ 
is chosen to have the same orientation throughout the bulk,
so that locally, at each brane, it is pointing in the same
direction and $q_{\alpha \beta} =q_{\alpha \beta}^{\;5D}-n_{\alpha}n_{\beta}$ if $q_{\alpha\beta}^{5D}$ is the
$5D$ metric. The proper distances between the branes are defined via
 \beq
 d_{1}(x)  = \int_{0}^{\tilde{y}}{e^{\phi(y',x)}dy'},
 \qquad
 d_{2}(x)  = \int_{\tilde{y}}^{y_0}{e^{\phi(y',x)}dy'}.
 \label{distances}
 \eeq

Our objective is to derive at first order the effective Einstein
equations on the third (i.e. middle) brane: this will be enough to
determine the effective Einstein equations on the other two
branes, as they are related to the Einstein equations on the third
brane by an appropriate conformal transformation, given the
specific form of the metric (\ref{RS metric}) which in turn
implies that the background metrics on the three branes can be
conformally transformed into each other.

In the perturbative scheme introduced in Ref.~\cite{AdS/CFT}, the
low-energy regime is defined as the regime in which the energy densities on the branes are negligible
with respect to the brane tension,
 \beq
\rho_i \ll |\sigma_i| \label{low regime},
 \eeq
or, taking into account (\ref{fine tuningRS}) and $8 \pi G_{(4)} =
\ka_5^2\ell^{-1}$ where $G_{(4)}$ is the effective $4D$ Newton's
constant \cite{Giappo},
 \beq
 \left(\frac{\ell}{L}\right)^2 \ll 1
 \eeq
where $\ell$ is the bulk curvature scale in the AdS slice and $L$
is the brane curvature scale. Therefore the background solution is
the vacuum spacetime, and perturbations are introduced as matter
is added on the branes. The parameter of expansion is given by
\cite{AdS/CFT}
 \beq
  \epsilon = \left(\frac{\ell}{L}\right)^2
 \eeq
and, accordingly, expansions of the extrinsic curvature
\beq
\label{defK}
K_{\alpha}^{\beta} \equiv
% q_{\alpha}^{\gamma} \nabla_{\gamma} n^{\beta} = 
\frac{1}{2} q^{\beta \sigma} \mathcal{L}_n q_{\alpha \sigma}
 = \frac12 q^{\beta \sigma} e^{-\phi}\partial_y q_{\alpha \sigma} \,,
\eeq
and the projected 5D Weyl tensor,
\beq
E_{\alpha}^{\beta}\equiv C^{\beta}_{\;\rho \alpha \sigma}n^{\rho}n^{\sigma}
\eeq
around the vacuum solutions can be considered as follows \cite{Low
energy}
\bea
 K_{\alpha}^{\beta} & = & K_{\alpha}^{\beta\, (0)} + \epsilon\, K_{\alpha}^{\beta\, (1)} +
  \epsilon^2\,K_{\alpha}^{\beta\, (2)} +\ldots \label{expansionI}\\
 E_{\alpha}^{\beta} & = & \epsilon \,E_ {\alpha}^{\beta\, (1)} +
  \epsilon^2\, E_{\alpha}^{\beta\, (2)}+ \ldots
\label{expansionII}
 \eea
where $E_{\alpha}^{\beta\,(0)} = 0$ for the Randall-Sundrum
brane-world. With the expansions (\ref{expansionI}) and (\ref{expansionII}), the evolution
equations for $K_{\alpha}^{\beta}$ and $E_{\alpha}^{\beta}$ can be
solved at different orders, and subsequentlty, so can the induced
Einstein equations on the brane \cite{AdS/CFT} \cite{Low energy}.
In particular the evolution equations in the bulk for
$K_{\alpha}^{\beta}$ and $E_{\alpha}^{\beta}$ are given by the Lie
derivatives along $n^{\alpha}$, as these describe the changes of
both the tensors along the integral curves of the normal vector
field $n^{\alpha}$.

At zero order for three branes, i.e. for the background solution,
the equations are no more complicated than in the two brane case.
The bulk solution between each pair of branes corresponds to a
region of anti-de Sitter, with curvature scales $\ell_1>\ell_2$,
and the Weyl tensor vanishes at this order. In each region the
extrinsic curvature tensor is constant but has two different
values in the two different slices of AdS spaces (giving rise to a
constraint on the tension of the bulk brane).

At first order, some further steps are required, the reason being mainly
that the third brane breaks the otherwise continuous evolution of the Weyl tensor in the bulk.
Briefly, the plan for deriving the first-order effective theory is
the following
\begin{itemize}
\item \textit{Junction conditions}\\
Write the junction conditions for each of the three branes. In
particular the third brane ($y=\tilde{y}$) is not at a fixed point
of the orbifold so that we have to include the jump suffered by
the extrinsic curvature tensor at the passage through brane III.
\item \textit{Evolution equations}\\
Write the evolution equations at first order (i.e. the Lie
derivatives at first order) for $E_{\a}^{\b}$ and $K_{\a}^{\b}$
which are the same as in the two brane scenario. However the
first-order solutions to these equations now hold only separately
in the two AdS regions of spacetime, so special attention is
required at brane III where in general the solutions are
discontinuous. In particular, from the Lie derivative of
$K_{\a}^{\b}(\tilde{y})$, we write the junction condition at the third middle brane
as a function of the source terms, the tensor
$E_{\a}^{\b}$ at both sides of brane III ($+$ and $-$ as sketched
in figure \ref{fig:brane_3_2}) and the kinetic terms associated
with the proper distances $d_1$ and $d_2$ (\ref{distances}).
\item \textit{Consistency of Einstein equations on brane III}\\
Impose the requirement of consistency of the induced Einstein equations on the
third brane. This leads to a junction condition for $E_{\a}^{\b}$ at this brane.
\item \textit{Weyl tensor}\\
Obtain a system of two independent equations in the two unknowns
$E_{\a}^{\b}(\tilde{y})^+$ and $E_{\a}^{\b}(\tilde{y})^-$, from
the previous steps. The system is then solved for
$E_{\a}^{\b}(\tilde{y})^+$ (analogously it could be solved for
$E_{\a}^{\b}(\tilde{y})^-$). We then obtain the Weyl tensor on one
side of brane III as a function of the sources and the kinetic
terms associated with both $d_1$ and $d_2$ (\ref{distances}).
\item \textit{Einstein equations}\\
Finally, substituting the expressions for $K_{\a}^{\b}(\tilde{y})^+$ and $E_{\a}^{\b}(\tilde{y})^+$ in equation
\cite{Low energy}
\beq
G_\alpha ^\beta  |_{\tilde y}^ +   =
  \left[ { - \frac{2}{{\ell_2 }}[K_{\a}^{\b}-\delta_{\a}^{\b}K]|_{\tilde y}^+
- E_\alpha ^\beta  |_{\tilde y}^ +  } \right] \label{G first order}
\eeq
enables us to obtain the Einstein equations on brane III.
\end{itemize}
We show that the effective gravity theory obtained at first-order
is a generalized Brans-Dicke theory with two scalar fields.

%%%%%%%%%%%%%%%%%%%%%%%%%%%%

\subsection{Junction conditions and background solution} \label{zerorder}

If the direction of the normal vector field $n^{\a}$ to a brane is chosen to be the same
throughout the bulk, the junction conditions read
\begin{itemize}
\item brane I \beq \left(K_{\a}^{\b} - \d_{\a}^{\b} K\right)\vert_{0}  =
-\frac{\ka_5^2}{2}\left(-\sigma_0 \d_{\a}^{\b}  +
T_{\a}^{\b}|_0 \right)\label{jc braneI}\eeq
\item brane III \beq \left[K_{\a}^{\b} - \d_{\a}^{\b} K\right]_{-}^{+} \vert_{\tilde{y}} =
-\ka_5^2\left(-\sigma_{\tilde{y}} \d_{\a}^{\b}+T_{\a}^{\b}|_{\tilde{y}} \right)\label{jc braneIII}\eeq
\item brane II \beq
\left(K_{\a}^{\b} - \d_{\a}^{\b} K\right)\vert_{y_0}  =
\frac{\ka_5^2}{2}\left(-\sigma_{y_0} \d_{\a}^{\b}  +
T_{\a}^{\b}|_{y_0} \right)\label{jc braneII}\eeq
 \end{itemize}
where the factor $1/2$ appears on the right-hand-side of
Eqs.~(\ref{jc braneI}) and~(\ref{jc braneII}) because we have
Z$_2$-symmetry at the boundary branes. We therefore assume the tension on brane I to be positive ($\sigma_0>0$) and
the tension on brane II to be negative ($\sigma_{y_0}<0$).

At the lowest order, matter is neglected and as in the two branes case the
only equation to solve is the equation for $K_{\a}^{\b}$, as
$E_{\a}^{\b}$ is taken to be zero at this order.
In the bulk and at this order, the Lie derivative for $K_{\a}^{\b}$ reads \cite{Low energy}
 \beq
 \mathcal{L}_n K_{\a}^{\b\,(0)} = e^{-\phi}\partial_y K_{\a}^{\b \,\,(0)} =
 \frac{1}{\ell^2}\d_{\a}^{\b}-  K_{\a}^{\gamma\,(0)}K_{\gamma}^{\b \,\,(0)}.
 \label{K zero order}
 \eeq
In each AdS slice equation (\ref{K zero order}) has solution \cite{Low energy}
 \beq
 K_{\a}^{\b\,(0)}= -\frac{1}{\ell}\d_{\a}^{\b}\label{sol K zero order}\,,
 \eeq
where $\ell$ is now either $\ell_1$ or $\ell_2$ depending on which
AdS slice is under consideration. Moreover, from 
%\beq
%K_{\alpha \beta} = \frac{1}{2} \mathcal{L}_n q_{\alpha \beta},
%\eeq
the definition of the extrinsic curvature (\ref{defK}), 
in each AdS region the metric at zero order reads \cite{Low energy}
 \beq
 q_{\a\b}^{(0)}(y,x) = e^{-2d(y,x)/\ell}h_{\a\b}(x) \label{metrics zero order}
 \eeq
where
 \beq
 \label{defd}
d \equiv \int_{\bar{y}}^y{e^{\phi(y',x)}dy'}
 \eeq
is the proper distance between $\bar{y}$, any fixed point on the
extra-coordinate axis, and $y$, both points being in the same AdS
region, and $h_{\a\b}(x)$ is a tensor field which does not depend
on the extra coordinate $y$ (but will in general depend on the
coordinates on some hypersurface orthogonal to the
extra-coordinate, and in particular on the coordinates on the
branes).

In our generalised Randall-Sundrum type scenario equations
(\ref{sol K zero order}), (\ref{jc braneI}) and (\ref{jc braneII})
imply
 \beq
K_{\a}^{\b} = \left\{\begin{array}{cc} -\frac{\ka_5^2}{6}\sigma_0 \delta_{\a}^{\b} & 0<y<\tilde{y},\\
\frac{\ka_5^2}{6}\sigma_{y_0} \delta_{\a}^{\b} & \tilde{y}<y<y_0 \end{array}\right.
 \label{K zero order 3brane}
 \eeq

Furthermore for the background solution we can write
\beq
\left[K_{\a}^{\b} - \d_{\a}^{\b} K\right]_{-}^{+} \vert_{\tilde{y}}=
\left[K_{\a}^{\b} - \d_{\a}^{\b} K\right]_{-}^{+} \vert_{y_0} - \left[K_{\a}^{\b} -
\d_{\a}^{\b} K\right]_{-}^{+} \vert_{0}\label{junction simple}
\eeq
and we then conclude, using solution (\ref{K zero order 3brane}), that there exits a fine
tuning condition constraining the tension on brane III, namely \cite{Gibbons}
\beq
\sigma_0 + \sigma_{y_0}  + 2\sigma_{\tilde{y}}  =  0\,. \label{Constraint background}
\eeq

Finally, again from solution (\ref{sol K zero order}), the condition (\ref{Constraint background})
can be expressed in terms of the AdS curvature scales $\ell_{1,2}$ as
\beq
\sigma_{\tilde{y}} = \frac{3}{\ka_5^2} \frac{(\ell_1 - \ell_2)}{\ell_1\,\,\ell_2}\,.
\eeq
where for simplicity we will assume $\sigma_{\tilde{y}}>0$ and hence
$\ell_1>\ell_2$.

%%%%%%%%%%%%%%%%%%%%%%%%%%%%%%%%%%%%%%%%%
\subsection{First-order effective theory} \label{firstorder}

At first order and in the bulk, the Lie derivatives for the Weyl tensor and the extrinsic curvature read
\cite{Low energy}
 \bea
\mathcal{L}_n E_{\a\b}^{(1)} & = & \frac{2}{\ell} E_{\a\b}^{(1)} \\
\mathcal{L}_nK_{\a}^{\b \,\,(1)} & = & - (D_{\a}D^{\b}\phi+D_{\a}\phi D^{\b}\phi)
 +  \frac{2}{\ell}K_{\a}^{\b \,\, (1)}- E_{\a}^{\b \,\,(1)}
 \eea
where $D^\alpha$ is the covariant derivative with respect to the
induced metric $q_{\mu\nu}(y)$.
These first-order evolution equations have solutions in each AdS
slice \cite{Low energy}
 \beq
E_{\a}^{\b \, \,(1)} (y,x)  =  e^{4\,d/\ell} E_{\a}^{\b \,\,(1)}(\bar{y},x)
\label{E_1}
 \eeq
where the zero-order induced metric (\ref{metrics zero order}) has
been used to raise the indices, and \cite{Low energy}
 \bea
K_{\a}^{\b \,\, (1)} (y,x) & = & e^{2\,d/\ell}K_{\a}^{\b \,\, (1)}(\bar{y},x)
- \frac{\ell}{2}\left[1-e^{-2\,d/\ell}\right]E_{\a}^{\b \,\,(1)}(y,x) \nonumber \\
& - & \left[D_{\a}D^{\b}d - \frac{1}{\ell}\left(D_{\a}d D^{\b}d
-\frac{1}{2}\d_{\a}^{\b}(Dd)^2\right)\right] \label{K_1}
 \eea
where again $d$ is the proper distance (\ref{defd}) between a
generic point $\bar{y}$ on the extra-coordinate axis and $y$.

In the following we drop the first-order superscripts, assuming
that all the quantities are evaluated at first order.

We define
 \beq
\bar{K}_\alpha ^\beta  |_{y} \equiv K_\alpha ^{\beta} (y,x) - \delta_\alpha ^\beta K (y,x).
 \eeq
{}From equation (\ref{K_1}), and integrating from brane I to brane
III, we have
 \bea
\bar{K}_\alpha ^\beta  |_{\tilde y}^ -  & = & e^{2d_1 /\ell_1 } \left( { - \frac{{\kappa _5^2 }}{2}T_\alpha ^\beta  |_0 }
\right) - \frac{{\ell_1 }}{2}\left( {1 - e^{ - 2d_1 /\ell_1 } } \right)
E_\alpha ^\beta  |_{\tilde y}^ -  \nonumber \\
& - &
\left( {D_\alpha  D^\beta   - \delta _\alpha ^\beta  D^2 } \right)d_1  + \frac{1}{{\ell_1 }}\left( {D_\alpha  d_1 D^\beta
d_1  + \frac{1}{2}\delta _\alpha ^\beta  (Dd_1 )^2 } \right)\label{K-}
\eea
where we have used equation (\ref{jc braneI}).
On the other side of brane III, but integrating back from brane
II, we have
 \bea
\bar{K}_\alpha ^\beta  |_{\tilde y}^ +  & = & e^{ - 2d_2 /\ell_2 } \left( {\frac{{\kappa _5^2 }}{2}
T_\alpha ^\beta  |_{y_0 } } \right) - \frac{{\ell_2 }}{2}\left( {1 - e^{2d_2 /\ell_2 } } \right)
E_\alpha ^\beta  |_{\tilde y}^ +  \nonumber \\
& + &
\left( {D_\alpha  D^\beta   - \delta _\alpha ^\beta  D^2 } \right)d_2  + \frac{1}{{\ell_2 }}
\left( {D_\alpha  d_2 D^\beta  d_2  + \frac{1}{2}\delta _\alpha ^\beta  (Dd_2 )^2 } \right)\label{K+}
 \eea
where this time we have used equation (\ref{jc braneII}). Note that
in Eqs.~(\ref{K-}) and~(\ref{K+}) $D^\alpha$ is the covariant
derivative induced on brane III.

Due to the junction conditions for $K_\alpha ^\beta$ on the two
sides of brane III, $E_{\a}^{\b}$ does not evolve continuously
from brane I to brane II, unlike the two brane scenario where
(\ref{E_1}) can be applied everywhere in the bulk.
In fact, continuity of the induced metric implies that the induced
Einstein tensor should also be continuous
 \beq
 G_\alpha ^\beta  |_{\tilde y}^ +   =  G_\alpha ^\beta  |_{\tilde y}^- \,.
 \eeq
At first order Eq.~(\ref{G first order}) thus gives a junction
condition for the projected Weyl tensor across the bulk brane
\beq
 E_\alpha ^\beta  |_{\tilde y}^ +   - E_\alpha ^\beta  |_{\tilde y}^ -    =
 - \frac{2}{{\ell_2 }}\bar{K}_\alpha ^\beta  |_{\tilde y}^ +   + \frac{2}{{\ell_1 }}
\bar{K}_\alpha ^\beta  |_{\tilde y}^ -
\label{E unique}
\eeq

In order to obtain an explicit expression for $E_\alpha ^\beta
|_{\tilde y}$ on either side of brane III, we extract from
equations (\ref{jc braneIII}), (\ref{K-}), (\ref{K+}) and (\ref{E
unique}) a set of two independent equations for the unknowns
$E_\alpha ^\beta  |_{\tilde y}^+$ and $E_\alpha ^\beta  |_{\tilde
y}^-$, which can then be solved. From (\ref{jc braneIII}),
(\ref{K-}) and (\ref{K+}) we obtain a first equation for $E_\alpha
^\beta  |_{\tilde y}^+$ and $E_\alpha ^\beta  |_{\tilde y}^-$,
 \bea
\frac{{\ell_2 }}{2}\,\mathcal{A}\,E_{\a} ^{\b}|_{\tilde{y}}^+
  - \frac{{\ell_1 }}{2}\,\mathcal{B}\,E_{\a}^{\b}|_{\tilde{y}}^-  & = & \frac{{\kappa_5^2}}{2}
\left(2 T_{\a}^{\b}|_{\tilde{y}} + e^{- 2d_2 /\ell_2 } T_{\a}^{\b}|_{y_0 }
+  e^{2d_1 /\ell_1 } T_{\a}^{\b}|_0 \right)\nonumber \\
& + & \left( {D_{\a}D^{\b} - \delta _{\a}^{\b}D^2}\right)(d_2  + d_1 ) \nonumber \\
+  \frac{1}{{\ell_2 }}\left({D_{\a}d_2D^{\b}d_2  + \frac{1}{2}\delta _{\a}^{\b}(Dd_2 )^2}\right)
& - & \frac{1}{{\ell_1 }}\left( {D_{\a}d_1 D^{\b}d_1 + \frac{1}{2}\delta _{\a}^{\b} (Dd_1 )^2 } \right)
\label{First combination}
\eea
where
\bea
\mathcal{A} & = & \left( {1 - e^{2d_2 /\ell_2 } } \right) \nonumber \\
\mathcal{B} & = & \left( {1 - e^{ - 2d_1 /\ell_1 } } \right)
\eea
The second independent equation for $E_\alpha ^\beta  |_{\tilde y}^+$ and
$E_\alpha ^\beta  |_{\tilde y}^-$ is directly obtained from (\ref{E unique}) and equations (\ref{K-}),
(\ref{K+})
\bea
 E_\alpha ^\beta  |_{\tilde y}^ +   - E_\alpha ^\beta  |_{\tilde y}^ -
 &  = & \left( {1 - e^{2d_2 /\ell_2 } } \right)E_\alpha ^\beta  |_{\tilde y}^+ \nonumber \\
&  - & \left( {1 - e^{ - 2d_1 /\ell_1 } }
\right)E_\alpha ^\beta  |_{\tilde y}^ -   - \frac{{\kappa _5^2 }}{{\ell_2 }}\left( {e^{ - 2d_2 /\ell_2 }
T_\alpha ^\beta  |_{y_0 } } \right) \nonumber \\
& - &
 \frac{{\kappa _5^2 }}{{\ell_1 }}\left( {e^{2d_1 /\ell_1 } T_\alpha ^\beta  |_0 } \right) -
2\left( {D_\alpha  D^\beta   - \delta _\alpha ^\beta  D^2 } \right)\left(\frac{{d_2 }}{{\ell_2 }} + \frac{{d_1 }}
{{\ell_1 }}\right)
\nonumber \\
&  - & 2\left( {D_\alpha  \frac{{d_2 }}{{\ell_2 }}D^\beta  \frac{{d_2 }}{{\ell_2 }} + \frac{1}{2}\delta _\alpha ^\beta
\left(D\frac{{d_2 }}{{\ell_2 }}\right)^2 } \right) \nonumber \\
 &  + & 2\left( {D_\alpha  \frac{{d_1 }}{{\ell_1 }}D^\beta  \frac{{d_1 }}{{\ell_1 }} + \frac{1}{2}\delta _\alpha ^\beta
\left(D\frac{{d_1 }}{{\ell_1 }}\right)^2 } \right)
\label{Second combination}
\eea

Equations~(\ref{First combination}) and~(\ref{Second combination})
can be solved, for instance, for $E_\alpha ^\beta  |_{\tilde y}^
+$ \cite{MPhil}, and the unique induced Einstein equation on the
bulk brane are then derived, using the obtained expression for
$E_{\a}^{\b}|_{\tilde{y}}^+$ in equation (\ref{G first order}).
After some rearrangements, we finally obtain
 \bea
G_\alpha ^\beta  |_{\tilde y}   & = & \frac{\ka_5^2}{\ell_3}\left[T_\alpha ^\beta  |_{\tilde{y}} +
\frac{e^{4d_1/\ell_1}}{2}T_\alpha ^\beta  |_{0}\right] +
\frac{\ka_5^2}{2{\ell}_3}e^{-4d_2/\ell_2}T_\alpha ^\beta  |_{y_0 } \nonumber \\
& + & \frac{e^{2d_1/\ell_1}}{\ell_3}[D_{\a}D^{\b}-\delta_{\a}^{\b}D^2]d_1
+\frac{e^{-2d_2/\ell_2}}{\ell_3}[D_{\a}D^{\b}-\delta_{\a}^{\b}D^2]d_2\nonumber \\
& - & \frac{e^{2d_1/\ell_1}}{\ell_1\ell_3}\left[D_{\a}d_1D^{\b}d_1 + \frac{1}{2}\delta_{\a}^{\b}D^2d_1\right]
\nonumber \\
& + & \frac{e^{-2d_2/\ell_2}}{\ell_2\ell_3}\left[D_{\a}d_2D^{\b}d_2 + \frac{1}{2}\delta_{\a}^{\b}D^2d_2\right]
\label{G 3brane}
\eea
where $\ell_3$ is defined as follows
\beq
\ell_3 \equiv \frac{\ell_1}{2}(e^{ 2d_1 /\ell_1}-1) + \frac{\ell_2}{2}(1-e^{ - 2d_2 /\ell_2})=
\ell_1e^{d_1/\ell_1}\sinh{\left(\frac{d_1}{\ell_1}\right)} + \ell_2e^{-d_2/\ell_2}\sinh{\left(\frac{d_2}{\ell_2}\right)}.
\label{elle3}
\eeq

We now show explicitly that the effective theory at first order is
indeed a generalised Brans-Dicke theory with two independent
scalar fields. We may define a first dimensionless scalar field
$\Phi$ to be
 \beq
\ell \Phi \equiv \ell_3
\label{defPhi}
 \eeq
where $\ell$ is an arbitrary unit of length and $\ell_3$ is given in (\ref{elle3}), 
so that the kinetic terms for $d_1$ and $d_2$ in (\ref{G 3brane})
can be rewritten as
\beq
\frac{1}{\Phi}[D_{\a}D^{\b} -\delta_{\a}^{\b}D^2]\Phi + \mathcal{C}
\eeq
where
\beq
\mathcal{C} = -\frac{3 e^{2d_1/\ell_1}}{\ell_1 \ell_3}
\left[D_{\a}d_1D^{\b}d_1 - \frac{1}{2}\delta_{\a}^{\b}D^2d_1\right]
+ \frac{3 e^{-2d_2/\ell_2}}{\ell_2 \ell_3}
\left[D_{\a}d_2D^{\b}d_2 - \frac{1}{2}\delta_{\a}^{\b}D^2d_2\right]\,\,.
\label{A}
\eeq

We can define a second scalar field, $\Psi$, which will have only
first order derivatives appearing in Eq.~(\ref{G 3brane}) and
should be as well a function of the two proper distances $d_1$ and
$d_2$ (\ref{distances}). In general we can write
$$
\Psi \equiv \Psi(u)
$$
where $u=u(d_1,d_2)$ and hence
\beq
\ell D_{\a}\Psi(u(d_1, d_2)) = \Psi_1 D_{\a}d_1 + \Psi_2 D_{\a}d_2
\eeq
where
$\Psi_i \equiv  \ell \Psi' \frac{\partial{u}}{\partial d_i}$ if $\Psi' = \frac{d\Psi}{d u}$.
Our objective is to be able to write $\mathcal{C}$ defined in equation (\ref{A}) as
\beq
\mathcal{C} = \frac{\omega(\Phi)}{\Phi^2}\left[\left(D_{\a}\Phi D^{\b}
 \Phi -\frac{1}{2}\delta_{\a}^{\b}\left(D\Phi\right)^2\right)
-\Gamma(\Phi)\left(D_{\a}\Psi D^{\b} \Psi -\frac{1}{2}\delta_{\a}^{\b}\left(D\Psi\right)^2\right)\right].
\label{request}
\eeq

Considering that in $\mathcal{C}$ there are no mixed terms
$D_{\a}d_1D^{\b}d_2$ or $D_{\a}d_2D^{\b}d_1$, we get a first
contraint on $\Gamma(\Phi)$ and $\Psi$, which reads
 \beq
e^{2d_1/\ell_1 -2d_2/\ell_2}  =  \Gamma \Psi_1\Psi_2
\label{constraint1}
 \eeq
where $\Phi_i \equiv  \ell (\partial{\Phi}/\partial d_i)$ and
$\Psi_i$ are defined as before.

A second constraint comes directly from the form of equation (\ref{request})
\bea
\ell^2\left[D_{\a}\Phi D^{\b} \Phi  -   \Gamma \left(D_{\a}\Psi D^{\b}\Psi \right) \right]
& = & \left[e^{4d_1/\ell_1} -\Gamma \Psi_1^2\right]D_{\a}d_1 D^{\b}d_1 \nonumber \\
& + &
\left[e^{-4d_2/\ell_2} -\Gamma \Psi_2^2\right]D_{\a}d_2 D^{\b}d_2
\label{first}
\eea
Comparing the ratio of the coefficients in equations (\ref{A}) and (\ref{first}) then implies
\beq
\frac{e^{4d_1/\ell_1} -\Gamma \Psi_1^2}{e^{-4d_2/\ell_2} -\Gamma \Psi_2^2} =
-\frac{\ell_2}{\ell_1}e^{2d_1/\ell_1 +2d_2/\ell_2}\hspace{0.25in}. \label{constraint2}
\eeq
Using (\ref{constraint1}) to eliminate $\Gamma$ in
(\ref{constraint2}), we obtain
 \beq
\left[1 -\left(\frac{\Psi_1}{\Psi_2}\right)e^{-2d_1/\ell_1 -2d_2/\ell_2}\right]
\left[\ell_1 -\ell_2 \left(\frac{\Psi_2}{\Psi_1}\right)\right] = 0\,\,.
\label{eq}
\eeq
The solutions of (\ref{eq}) are
 \beq
\frac{\Psi_1}{\Psi_2}  =  e^{2d_1/\ell_1 +2d_2/\ell_2}; \hspace{1in} \frac{\Psi_1}{\Psi_2}  =  \frac{\ell_2}{\ell_1}.
\eeq

The first solution corresponds to the limit in which $\Phi = \Psi$ and both coefficients on the right hand side
of (\ref{first}) vanish. Choosing the second solution of (\ref{eq}) and using the defintion of $\Psi_i$, we obtain
\beq
u = \frac{d_1}{\ell_1} + \frac{d_2}{\ell_2}\label{u}\,\,.
\eeq
{}From the first constraint (\ref{constraint1}) and using $\Psi_i
= \Psi' \ell(\ell_i)^{-1}$ we obtain a differential equation for
$\Psi(u)$ which reads
 \beq
\Psi'  =
\frac{\sqrt{\ell_1\ell_2}}{\ell\sqrt{\Gamma}}e^u \frac{2\ell\Phi +\ell_1 -\ell_2}{\ell_1 e^{2u} -\ell_2}.
\label{diff Psi}
\eeq
where we have used (\ref{defPhi}).

In order for $\Psi$ to be only a function of $u$, and not of
$\Phi$, as the scalar fields should be independent, we require
that in equation (\ref{diff Psi})
 \beq
  \ell^2 \Gamma(\Phi) =
 \left(\ell\Phi +\frac{\ell_1}{2} -\frac{\ell_2}{2}\right)^2\,\,.
\label{gamma}
 \eeq
Equation~(\ref{diff Psi}) then becomes
 \beq \Psi' =
\frac{2\sqrt{\ell_1\ell_2}e^u}{\ell_1e^{2u} -\ell_2}
 \eeq
and integrating we obtain
 \beq
  \Psi (u) =
\ln{\left|\left(\frac{z-1}{z+1}\right)\right|}\,\,\,\,\,\,\mbox{where}\,\,\,\,\,\,z
= \sqrt{\frac{\ell_1}{\ell_2}}e^u \label{defPsi} \eeq so that \beq
e^u =
\sqrt{\frac{\ell_2}{\ell_1}}\coth\left(\frac{\Psi}{2}\right)\,\,.
 \label{eu}
 \eeq

Imposing now (\ref{request}), from both (\ref{constraint1}) and (\ref{constraint2}), we obtain
 \beq
\omega(\Phi) = -\frac{3}{2}\left(\frac{\ell\Phi}{\ell\Phi + \left(\frac{\ell_1}{2} -\frac{\ell_2}{2}\right)}\right)\,\,.
\label{omega3}
\eeq

Finally, the Einstein equations on brane III (\ref{G 3brane}) can be written as
\bea
G_\alpha ^\beta  |_{\tilde y}   & = & \frac{\ka_5^2}{\ell \Phi}T_\alpha ^\beta  |_{\tilde{y}} +
\frac{\ka_5^2}{\Phi}\left(\frac{2\ell\Gamma}{\ell_1^2}\right)
\cosh^4\left(\frac{\Psi}{2}\right)T_\alpha ^\beta  |_{0} \nonumber \\
& + &
\frac{\ka_5^2}{\Phi}\left(\frac{2\ell\Gamma}{\ell_2^2}\right)\sinh^4\left(\frac{\Psi}{2}\right)
T_\alpha ^\beta  |_{y_0 }
+ \frac{1}{\Phi}[D_{\a}D^{\b} -\delta_{\a}^{\b}D^2]\Phi \nonumber \\
& + & \frac{\omega(\Phi)}{\Phi^2}\left[D_{\a}\Phi D^{\b} \Phi -\frac{1}{2}\delta_{\a}^{\b}\left(D\Phi\right)^2
\right] \nonumber \\
& - & \frac{\omega(\Phi)}{\Phi^2}\Gamma(\Phi)
\left[D_{\a}\Psi D^{\b} \Psi -\frac{1}{2}\delta_{\a}^{\b}\left(D\Psi\right)^2\right]\,\,.
\label{Finale}
 \eea
where the scalar fields $\Psi$ and $\Psi$ are defined as in
Eqs.~(\ref{defPhi}) and (\ref{defPsi}), and $\Gamma(\Phi)$ and
$\omega(\Phi)$ are defined in Eqs.~(\ref{gamma}) and
(\ref{omega3}).

We can conclude from (\ref{Finale}) that at low energy the effective theory on brane III
is a generalized Brans-Dicke theory with two scalar fields, namely $\Psi$ and $\Phi$.
This conclusion is a generalization of the result obtained in \cite{AdS/CFT} and \cite{Low energy} for a
two brane system.
Note that although we find the Brans-Dicke parameter on the bulk
brane, $\omega(\Phi)$, is negative, it is greater than $-3/2$ and
thus there is no instability. This is most clearly seen when one
writes the effective action in terms of scalar fields minimally
coupled to the scalar curvature, i.e., in the Einstein frame. Also
$\omega(\Phi)\Gamma(\Phi)$ is negative and thus the kinetic terms
for the second field $\Psi$ are also are of the correct sign and
again there is no instability.

%%%%%%%%%%%%%%%%%%%%%%%%%%%%%%%%%%%%%%

\subsection{The Einstein frame} \label{Eframe}

In this last subsection, we show that the theory is, as expected for Brans-Dicke
theories, conformally equivalent to Einstein gravity with two scalar fields minimally coupled with
respect to the metric.

The effective action which yields the field equations
(\ref{Finale}), when written in terms of the induced metric on
brane III, reads
 \bea
S_{\tilde{y}} & = & \frac{\ell}{2\kappa_5^2} \int{d^4x\sqrt{-g}\left[\Phi\,R - \frac{\omega(\Phi)}{\Phi}\left[\left(\nabla\Phi\right)^2
-\Gamma(\Phi)\left(\nabla\Psi\right)^2\right]\right]} \nonumber \\
& + & \int{d^4x \sqrt{-g}\left[\mathcal{L}({\tilde{y}}) + \mathcal{F}_0(\Phi, \Psi)\mathcal{L}(0) +
\mathcal{F}_{y_0}(\Phi, \Psi)\mathcal{L}(y_0) \right]} \label{effective_action}
 \,,
 \eea
where
 \beq
\mathcal{F}_0  =  2\frac{\ell^2 \Gamma}{\ell_1^2}\cosh^4\left(\frac{\Psi}{2}\right)\hspace{0.025in}; \hspace{1in}
\mathcal{F}_{y_0}  =   2\frac{\ell^2 \Gamma}{\ell_2^2}\sinh^4\left(\frac{\Psi}{2}\right). \label{Fs}
 \eeq
Performing a conformal transformation $\tilde{g}_{\alpha\beta} = \Omega^2g_{\alpha\beta}$, the effective
action on brane III (\ref{effective_action}) becomes
 \bea
\tilde{S}_{\tilde{y}} & = & \frac{1}{2\kappa_4^2}\int{d^4x\sqrt{-\tilde{g}}\,\left[
\left[\tilde{R} -
\frac{3}{2}\left(\frac{\tilde{\nabla}\Phi}{\Phi}\right)^2\right] -
\omega(\Phi)\left[\left(\frac{\tilde{\nabla}\Phi}{\Phi}\right)^2 - \Gamma(\Phi)\left(\frac{\tilde{\nabla}\Psi}{\Phi}\right)^2\right]
\right]} \nonumber \\
& + & \frac{\ka_5^4}{\ell^2\ka_4^4}
\int{d^4x\sqrt{-\tilde{g}}\,\,\Phi^{-2}\,\left[\mathcal{L}_{\tilde{y}} + \mathcal{F}_0(\Phi, \Psi)\mathcal{L}(0)
+ \mathcal{F}_{y_0}(\Phi, \Psi)\mathcal{L}(y_0)\right]} \label{conformal action2}
 \eea
where we chose
 \beq
\Phi \,\,\Omega^{-2}  = \ka_5^2\left(\ell\ka_4^2\right)^{-1}
 \eeq
and the Ricci scalar is now the Ricci scalar with respect to the
conformally transformed metric. The constant $\ka_4$ is an
arbitrary constant which is related to the effective $4D$ Newton's
constant $G_{(4)}$ via $8 \pi G_{(4)}= \ka_4^2$.

Defining $\gamma=2\ell\,(\ell_1-\ell_2)^{-1}$ (which, given our choice $\ell_1>\ell_2$, is a positive constant)
and taking into account (\ref{gamma}) (\ref{omega3}), action (\ref{conformal action2}) becomes
 \bea
\tilde{S}_{\tilde{y}} & = & \frac{1}{2\kappa_4^2}\int{d^4x\sqrt{-\tilde{g}}\,
\left[\tilde{R} -
\frac{3}{2}\left[\frac{1}{(\gamma\Phi+1)}\left(\frac{\tilde{\nabla}\Phi}{\Phi}\right)^2 +
\tilde{\Gamma}(\Phi)\left(\tilde{\nabla}\Psi\right)^2\right]\right]} \nonumber \\
& + &
\frac{\ka_5^4}{\ell^2\ka_4^4}\int{d^4x\sqrt{-\tilde{g}}\,\,\Phi^{-2}
\,\left[\mathcal{L}_{\tilde{y}} + \mathcal{F}_0(\Phi, \Psi)\mathcal{L}(0)
+ \mathcal{F}_{y_0}(\Phi, \Psi)\mathcal{L}(y_0)\right]} \label{conformal action3}
 \eea
where
$\tilde{\Gamma}(\Phi)=\left(\gamma\Phi+1\right)(\gamma\Phi)^{-1}$.
We now define
 \beq
(\tilde{\nabla}\eta)^2 = \frac{3}{2} \left[\frac{1}{\gamma\Phi+1}\right]
\left(\frac{\tilde{\nabla}\Phi}{\Phi}\right)^2
\label{eta}
 \eeq
from which we get, as in the two brane case,
 \beq
\coth{\left(\frac{\eta}{\sqrt{6}}\right)}=\sqrt{\gamma\Phi +1}\label{eta2}.
 \eeq

%Note that, since $\Phi$ is positive, the solution
%$\sqrt{(\gamma\Phi+1)} = \tanh{\left(\eta/\sqrt{6}\right)}$ could
%not work as $|\tanh{x}| < 1 \, \, \forall x$.
Finally given
(\ref{eta2}) we have
 \bea
\tilde{S}_{\tilde{y}} & = & \frac{1}{2\kappa_4^2}\int{d^4x\sqrt{-\tilde{g}}\,
\left[\tilde{R} - \left(\tilde{\nabla}\eta\right)^2 -\frac{3}{2}\cosh^2{\left(\frac{\eta}{\sqrt{6}}\right)}
\left(\tilde{\nabla}\Psi\right)^2\right]} \nonumber \\
%& + &
%\frac{\ka_5^4\gamma^2}{\ell^2\ka_4^4}\int{d^4x\sqrt{-\tilde{g}}\,\,
%\sinh^4{\left(\frac{\eta}{\sqrt{6}}\right)}
%\left[\mathcal{L}_{\tilde{y}} + \mathcal{{G}}_0(\eta, \Psi)\mathcal{L}(0)
%+ \mathcal{{G}}_{y_0}(\eta, \Psi)\mathcal{L}(y_0)\right]} \label{conformal action finale}
% \eea
%where now the $\mathcal{{G}}_0$ and $\mathcal{{G}}_{y_0}$
%are functions of the new scalar field $\eta$ and the old $\Psi$. The source
%terms in (\ref{conformal action finale}) could be further arranged so to become
% \beq
&&+
\frac{\ka_5^4\gamma^2}{\ell^2\ka_4^4}\int{d^4x\sqrt{-\tilde{g}}\,\,
\sinh^4{\left(\frac{\eta}{\sqrt{6}}\right)}
\mathcal{L}_{\tilde{y}}} \nonumber\\
&& + \frac{\ka_5^4\gamma^2}{\ka_4^4}\int{d^4x\sqrt{-\tilde{g}}\,\,
 2\cosh^4{\left(\frac{\eta}{\sqrt{6}}\right)}\left[\frac{
\cosh^4{\left(\frac{\Psi}{2}\right)}}{\ell_1^2}\mathcal{L}(0) +
\frac{
\sinh^4{\left(\frac{\Psi}{2}\right)}}{\ell_2^2}\mathcal{L}(y_0)\right]}
 \,.
 \eea
%taking into account (\ref{Fs}).

\section{Conclusions and Discussions} \label{conclusions}

In this paper we derived the low-energy effective theory for a
generalized Randall-Sundrum scenario, with three $4D$ branes
embedded in a $5D$ AdS bulk.  Two of the branes are located at the fixed points of
the orbifold, but the third brane can be located anywhere in between.
By construction, the metrics on the three branes are all connected by an appropriate conformal transformation.
It is therefore enough to derive the $4D$ effective Einstein equations on the third brane, as the
effective theory on the other branes can then be obtained performing a conformal transformation.

We followed the covariant approach adopted in Ref.~\cite{Low energy} to derive the low-energy effective theory
for a two brane Randall-Sundrum system. We considered an expansion of the extrinsic curvature and of
the projected Weyl tensor, where the
expansion parameter is the ratio of the energy density on the brane to the vacuum energy density as in
Refs.~\cite{AdS/CFT,Low energy}. In each separate region of the AdS bulk,
the evolution equations for the extrinsic
curvature tensor and for the projected Weyl tensor are, at each order, the same as in the two brane scenario and
solutions to the first-order equations can be found separately in each AdS slice.

The presence of the third brane obliges the evolution of the projected Weyl tensor to be discontinous.
{}From the requirement of consistency of the Einstein equations and from the junction conditions for the extrinsic
curvature tensor at the third brane we obtain a junction condition
for the projected Weyl tensor in terms of the extrinsic curvatures on
both sides of the branes and the sources on both branes.
Once an expression for the Weyl tensor as a function of the sources on the branes was obtained,
we have finally derived the first order $4D$ effective Einstein
equations on the third brane. The resulting theory is
a generalized Brans-Dicke theory with two independent scalar fields.
The appearance of two independent scalar fields is not surprising as the three brane scenario
is characterized by two natural scalar degrees of freedom:
the overall size of the orbifold and the position of the third brane.
A non-minimal coupling of the fields is found with respect to matter on the other two branes.
We have then showed that the effective theory is conformally equivalent to Einstein gravity plus two scalar fields
minimally coupled with the geometry.

We can conclude that the interpretation of the radion field in the two brane scenario can be
generalized to a three-brane scenario in which there exists an additional scalar degree of freedom. In the
two brane case the realization at first order of the non-local Einstein gravity,
with the generalized dark radiation term, as a local effective theory is described by the radion field
which appears in the equations through its derivatives.
In the three brane case, as in the case where a scalar bulk field is living in the bulk \cite{Palma, Dilaton},
two scalar fields both contribute to the realization of the (local) effective theory on the brane.

A moving bulk brane of the sort described here was discussed as a simple realisation of the original ekpyrotic
scenario \cite{Turok} where the collision of the bulk brane with a Minkowski boundary brane was interpreted as
initiating a hot big bang cosmology on the brane. (A moving bulk brane has also been studied in M-theory effective
action \cite{GrayLukas}.) Unlike the collision of two boundary branes in the later cyclic model \cite{Steinhardt},
the bulk spacetime does not disappear at the collision of a bulk brane with the boundary and hence the outgoing
state is completely determined by the incoming state~\cite{Collision of branes}.
If the boundary brane tension does not obey the RS fine-tuning (\ref{fine tuningRS}) then it may be inflating before
the collision and the possibility that a collision of the bulk brane with the boundary may trigger the end of
inflation was studied in Ref.~\cite{KSW}, using the effective action derived in this paper. It remains to be seen
whether the techniques described in this paper might be suitable for deriving a low-energy effective action capable
of incorporating the gravitational back-reaction of moving branes in flux-compactification scenarios, see, e.g.,
Refs.~\cite{KKKK,Baumann}.

\section*{Acknowledgements}

LSR acknowledges support from the Swiss National Science Foundation, grant number 205320-103833/1.
DW is supported in part by PPARC grant number PPA/G/S/2002/00576.

%%%%%%%%%%%%%%%%%%%%
%%%%%%%%%%%%%%%%%%%
%%%%%%%%%%%%%%%%%%%


\begin{thebibliography}{99}
%%%%%%%%%%%%%%%%%%%%%%%%%M THEORY AND TUROK%%%%%%%%%%%%%%%%%%%%%%
%%%%%%%%%%%%%%%%%%%%%%%%%%%%%%%%%%%%%%%%%%%%%%%%%%%%%%%%%%%%%%
\bibitem{M theory}
  A.~Lukas, B.~A.~Ovrut and D.~Waldram, %``Cosmological solutions of Horava-Witten theory'',
  Phys.\ Rev.\ D {\bf 60}, 086001 (1999)
  [arXiv:hep-th/9806022]; A.~Lukas, B.~A.~Ovrut and D.~Waldram, %``Heterotic M theory in five dimensions'',
  Nucl.\ Phys.\ B {\bf 552}, 246 (1999)
  [arXiv:hep-th/9806051];
  M.~Brandle, A.~Lukas and B.~A.~Ovrut, %``Heterotic M-Theory Cosmology in Four and Five Dimensions'',
  Phys.\ Rev.\ D {\bf 63}, 026003 (2000)
  [arXiv:hep-th/0003256].

\bibitem{Randall SundrumI}
  L.~Randall and R.~Sundrum, %``A Large Mass Hierarchy from a Small Extra Dimension'',
  Phys.\ Rev.\ Lett.\  {\bf 83}, 3370 (1999)
  [arXiv:hep-ph/9905221].

\bibitem{Randall SundrumII}
  L.~Randall and R.~Sundrum, %``An Alternative to Compactification'',
  Phys.\ Rev.\ Lett.\  {\bf 83}, 4690 (1999)
  [arXiv:hep-th/9906064].
  %\bibitem{Arkani}
  %N.~Arkani-Hamed, S.~Dimopoulos and G.~Dvali, ``The Hierarchy Problem and New Dimensions at a Millimeter'', Phys.\
  %Lett.\  B {\bf 429} (1998) 263.

\bibitem{Giappo}
  T.~Shiromizu, K.~Maeda and M.~Sasaki, %``The Einstein Equation on the 3-Brane World'',
  Phys.\ Rev.\ D {\bf 62}, 024012 (2000)
  [arXiv:gr-qc/9910076].

%\bibitem{BDL}
%  P.~Binetruy, C.~Deffayet and D.~Langlois, %``Non-conventional cosmology from a brane-universe'',
%  Nucl.\ Phys.\ B {\bf 565}, 269 (2000). ARCHIVE

%\bibitem{binetruy2}
%  P.~Binetruy, C.~Deffayet, U.~Ellwanger and D.~Langlois, %``Brane cosmological evolution in a bulk with
  %cosmological constant,''
%  Phys.\ Lett.\ B {\bf 477}, 285 (2000). ARCHIVE

%\bibitem{cosmo RS}
%  C.~Cs\'aki, M.~Graesser, C.~Kolda and J.~Terning,%``Cosmology of One Extra Dimension with Localized Gravity'',
%  Phys.\ Lett.\ B {\bf 462}, 34 (1999). ARCHIVE

\bibitem{Maartens}
  R.~Maartens, %``Cosmological dynamics on the brane'',
  Phys.\ Rev.\ D {\bf 62}, 084023 (2000)
  [arXiv:hep-th/0004166]; R.~Maartens, %``Geometry and dynamics of the brane world'',
  arXiv:gr-qc/0101059; R.~Maartens, %``Brane-world gravity'',
  Living Rev.\ Relativity  {\bf 7}, 7 (2004)
  [arXiv:gr-qc/0312059].
  %\textit{Reference Frames and Gravitomagnetism} J.~Pascual-Sanchez {\it et al.}, World Scientific 2001.
  
%%%%%%%%%%%%%%%%%%%%%%%%%%%%%%%%%%%%%%%%%%%%%%%
%%%%%%%%%%LOW ENERGY%%%%%%%%%%%%%%%%%%%%%%%%%%%%%%
%%%%%%%%%%%%%%%%%%%%%%%%%%%%%%%%%%%%%%%%%%%
\bibitem{AdS/CFT}
  S.~Kanno and J.~Soda, %``Brane World Effective Action at Low Energies and AdS/CFT Correspondence''
  Phys.\ Rev.\ D {\bf 66}, 043526 (2002)
  [arXiv:hep-th/0205188]; S.~Kanno and J.~Soda, %``Radion and holographic brane gravity'',
  Phys.\ Rev.\ D {\bf 66}, 083506 (2002)
  [arXiv:hep-th/0207029].

\bibitem{Wiseman}
  T.~Wiseman,
%``Strong brane gravity and the radion at low energies'',
  Class.\ Quant.\ Grav.\ {\bf 19}, 3083 (2002) [arXiv:hep-th/0201127].

\bibitem{Low energy}
  T.~Shiromizu and K.~Koyama, %``Low energy effective theory for two branes system:
  %covariant curvature formulation'',
  Phys.\ Rev.\ D {\bf 67}, 084022 (2003) [arXiv:hep-th/0210066].

\bibitem{Low energyII}
  S.~Kanno and J.~Soda, %``Braneworld Effective Action at Low Energies'',
  Astrophys.\ Space Sci.\  {\bf 283}, 481 (2003)
  [arXiv:gr-qc/0209087].

\bibitem{Turok}
  J.~Khoury, B.~A.~Ovrut, P.~J.~Steinhardt and N.~Turok, %``The ekpyrotic universe: Colliding branes and the origin of the hot big bang,''
  Phys.\ Rev.\ D {\bf 64}, 123522 (2001)
  [arXiv:hep-th/0103239]; A.~J.~Tolley, N.~Turok and P.~J.~Steinhardt, %``Cosmological perturbations in a big crunch / big bang space-time,''
  Phys.\ Rev.\ D {\bf 69}, 106005 (2004) [arXiv:hep-th/0306109].

\bibitem{Born again}
  S.~Kanno, M.~Sasaki and J.~Soda, %``Born-again brane world'',
  Prog.\ Theor.\ Phys.\ {\bf 109}, 357 (2003)
  [arXiv:hep-th/0210250].

\bibitem{MPhil}
  L.~Cotta-Ramusino, MPhil thesis, \textit{Low energy effective theory for brane world models}, Portsmouth University, 2004.

\bibitem{KKKK}
  K.~Koyama and K.~Koyama, %``Gravitational backreaction of anti-D branes in the warped
  %compactification,''
  Class.\ Quant.\ Grav.\  {\bf 22}, 3431 (2005)
  [arXiv:hep-th/0505256].
  %%CITATION = HEP-TH 0505256;%%

\bibitem{KSW}
  S.~Kanno, J.~Soda and D.~Wands, %``Braneworld flux inflation,''
  JCAP {\bf 0508}, 002 (2005)
  [arXiv:hep-th/0506167].
  %%CITATION = HEP-TH 0506167;%%

%\cite{deRham:2005pf}
\bibitem{deRham}
  C.~de Rham, S.~Fujii, T.~Shiromizu and H.~Yoshino, %``High-energy effective theory for a bulk brane,''
  Phys.\ Rev.\ D {\bf 72}, 123522 (2005)
  [arXiv:hep-th/0509194].
  %%CITATION = HEP-TH 0509194;%%

%%%%%%%%%%%%%%%%%%%%%%%%%%%%%%%%%%%%%DILATON AND MULTIBRANE SCENARIOS%%%%%%%%%%%%%%
%%%%%%%%%%%%%%%%%%%%%%%%%%%%%%%%%%%%%%%%%%%%%%%%%%%%%%%%%%%%%%%%%%%%%%%%%%%%%%%%%%%
\bibitem{Tre brane}
  I.~I.~Kogan, S.~Mouslopoulos, A.~Papazoglou and L.~Pilo, %``Radion in multibrane world''
  Nucl.\ Phys.\ B {\bf 625}, 179 (2002)
  [arXiv:hep-th/0105255].

\bibitem{Collision of branes}
  D.~Langlois, K.~Maeda and D.~Wands, %``Conservation laws for collisions of branes (or shells) in general relativity'',
  Phys.\ Rev.\ Lett.\ {\bf 88}, 181301 (2002)
  [arXiv:gr-qc/0111013].

\bibitem{Gibbons}
  G.~W.~Gibbons, R.~Kallosh and A.~D.~Linde,
   %``Brane world sum rules,''
  JHEP {\bf 0101}, 022 (2001)
  [arXiv:hep-th/0011225].
  %%CITATION = HEP-TH 0011225;%%

\bibitem{Palma}
  G.~A.~Palma and A.~C.~Davis, %``Low energy branes, effective theory and cosmology'',
  Phys.\ Rev.\ D {\bf 70}, 064021 (2004)
  [arXiv:hep-th/0406091];  G.~A.~Palma and A.~C.~Davis,  %``Moduli-space approximation for BPS brane worlds'',
  Phys.\ Rev.\ D {\bf 70}, 106003 (2004)
  [arXiv:hep-th/0407036].

\bibitem{Dilaton}
  S.~Kanno and J.~Soda, %  ``Low energy effective action for dilatonic braneworld: A formalism for
  %inflationary braneworld,''
  Gen.\ Rel.\ Grav.\  {\bf 36}, 689 (2004)
  [arXiv:hep-th/0303203].
  %%CITATION = HEP-TH 0303203;%%

\bibitem{GrayLukas}
  E.~J.~Copeland, J.~Gray and A.~Lukas, %``Moving five-branes in low-energy heterotic M-theory,''
  Phys.\ Rev.\ D {\bf 64}, 126003 (2001)
  [arXiv:hep-th/0106285];
  %%CITATION = HEP-TH 0106285;%%
  E.~J.~Copeland, J.~Gray, A.~Lukas and D.~Skinner, %``Five-dimensional moving brane solutions with four-dimensional limiting
  %behaviour,''
  Phys.\ Rev.\ D {\bf 66}, 124007 (2002)
  [arXiv:hep-th/0207281].
  %%CITATION = HEP-TH 0207281;%%

\bibitem{Steinhardt}
  P.~J.~Steinhardt and N.~Turok, %``A cyclic model of the universe,''
  arXiv:hep-th/0111030.
  %%CITATION = HEP-TH 0111030;%%

\bibitem{Baumann}
  D.~Baumann, A.~Dymarsky, I.~R.~Klebanov, J.~Maldacena, L.~McAllister and A.~Murugan, %``On D3-brane potentials in compactifications with fluxes and wrapped
  %D-branes,''
  arXiv:hep-th/0607050.
  %%CITATION = HEP-TH 0607050;%%



\end{thebibliography}
\end{document}